\documentclass[aps,prd,twocolumn]{revtex4}
\usepackage{graphicx}
\usepackage{dcolumn}

\def\beq{\begin{equation}}
\def\eeq{\end{equation}}
\def\bea{\begin{eqnarray}}
\def\eea{\end{eqnarray}}

\def\mpc{{\rm ~Mpc}}
\def\bwt{\begin{widetext}}
\def\ewt{\end{widetext}}

\usepackage{hyperref}
\nocite{*}

\begin{document}

\title{How well can (renormalized) perturbation theory predict dark matter clustering properties?}
\author{Niayesh Afshordi}\email{nafshordi@cfa.harvard.edu}\affiliation{Institute for
  Theory and Computation,
Harvard-Smithsonian Center for Astrophysics, MS-51, 60 Garden
Street, Cambridge, MA 02138}
\date{\today}
\preprint{hep-th/0610336}
\begin{abstract}
There has been some recent activity in trying to understand the dark
matter clustering properties in the quasilinear regime, through
resummation of perturbative terms, otherwise known as the
renormalized perturbation theory \cite{Crocce:2005xy}, or the
renormalization group method \cite{McDonald:2006hf}. While it is not
always clear why such methods should work so well, there is no
reason for them to capture non-perturbative events such as
shell-crossing. In order to estimate the magnitude of
non-perturbative effects, we introduce a (hypothetical) model of
{\it sticky dark matter}, which only differs from collisionless dark
matter in the shell-crossing regime. This enables us to show that
the level of non-perturbative effects in the dark matter power
spectrum at $k \sim 0.1 \mpc^{-1}$, which is relevant for baryonic
acoustic oscillations, is about a percent, but rises to order unity
at $k \sim 1 \mpc^{-1}$.

\end{abstract}
\maketitle

In the era of precision cosmology, an accurate understanding of dark
matter clustering properties is an essential ingredient of relating
observations of clustering in galaxy or weak lensing maps to the
fundamental properties of our Universe, such as its geometry or
linear growth history. This need is underlined by the recent
discovery of baryonic acoustic oscillations in the power spectrum of
galaxies \cite{Eisenstein:2005su,Cole:2005sx}, and the prospects of
its application as a cosmic standard ruler (e.g. \cite{Seo:2003pu}).
While semi-analytic methods, most famously within the context of the
halo model \cite{Seljak:2000gq,Cooray:2002di,Berlind:2002rn}, have
been successfully used to model both galaxy and dark matter
clustering properties, they suffer from their phenomenological
nature, which inevitably leads to a plethora of parameters that need
to be calibrated using numerical simulations or observations.
Moreover, the number of these parameters is expected to increase
through more in-depth studies, which will certainly lead to the
surfacing of more subtle effects (e.g.
\cite{Gao:2005ca,Wechsler:2005gb}). Thus, the ultimate dilemma in
the era of precision cosmology will become if we are probing the
fundamental properties of our universe, or rather further
constraining the intricate phenomenology of non-linear gravitational
gastrophysics.

Numerical simulations go a long way in elucidating the complicated
process of gravitational collapse of dark matter and baryonic gas.
However, they still suffer from our poor understanding of star and
galaxy formation, and its feedback on the surrounding intergalactic
medium. Moreover, they may also suffer from finite resolution and
box size effects, as well as potential ill-understood numerical
artifacts that can limit the accuracy of numerical studies.

The latter has motivated analytic studies, which are most
systematically done in the context of perturbation theory (see
\cite{Bernardeau:2001qr} for a review). However, consecutive terms
in a perturbative series become comparable in the quasilinear
regime, when overdensities become of order unity, signaling the
breakdown of the perturbative expansion. Inspired by the use of
renormalization methods in quantum field theory, some studies
\cite{Crocce:2005xy,McDonald:2006hf} have introduced renormalization
or resummation technics that, in principle, capture the dominant
terms in the perturbative series, and can yield an accurate result,
even when the perturbation theory breaks down.

While the theoretical validity of the technics advocated in these
studies is not transparent to this author, it is possible to put a
limit on the degree of accuracy that any method, based on
perturbation theory, can achieve. This limit comes from the
non-perturbative effects involved in the gravitational collapse
process, and is the subject of this note.

\begin{figure*}
\includegraphics[width=0.45\linewidth]{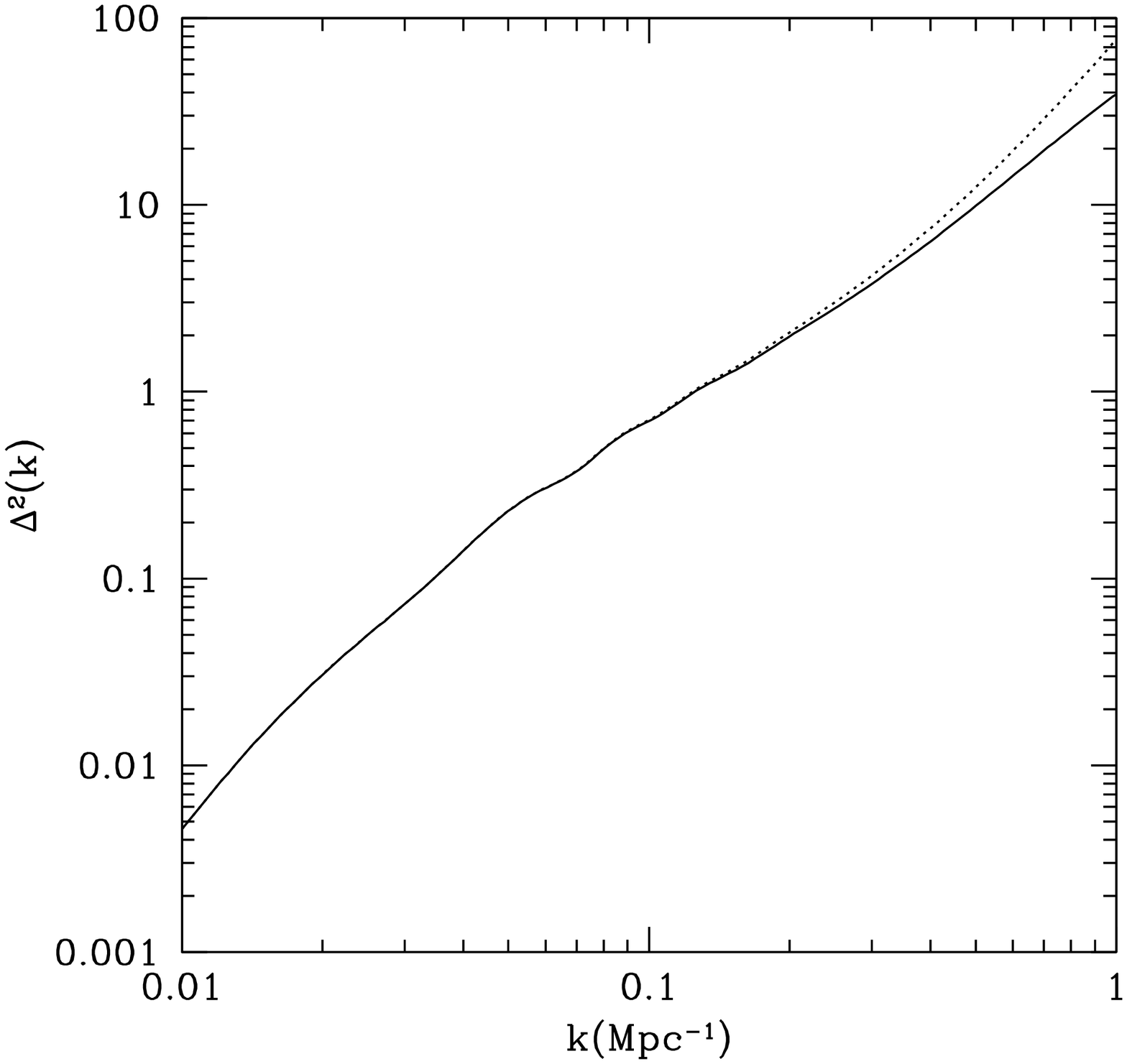}
\includegraphics[width=0.45\linewidth]{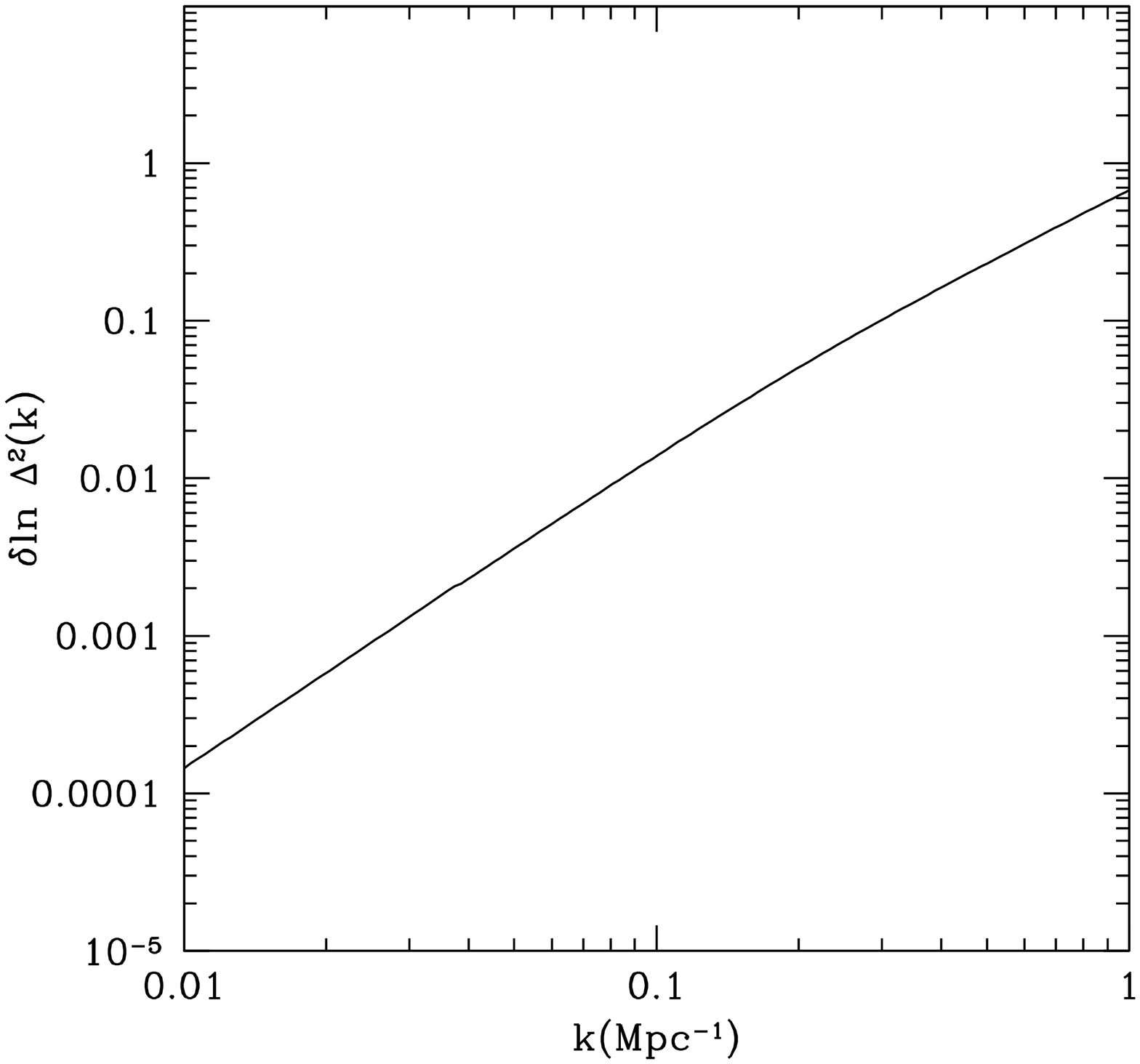}
\caption{\label{deltak} {\it Left}: Halo model dark matter power
spectrum for collisionless (solid) and sticky (dotted) dark matter.
{\it Right}: Relative difference between power spectra of
collisionless and sticky dark matter models. This shows the level of
non-perturbative effects in the dark matter power spectrum. }
\end{figure*}

The cosmological perturbation theory for cold dark matter
\cite{Bernardeau:2001qr} is based on the pressureless Euler
equation, where the fluid starts from a nearly homogenous expanding
initial conditions, and then evolves under its own gravity. The
equations break down at shell crossing, which is when perturbations
grow to the limit that multiple streams cross each other, and thus
the assumption of zero pressure fails.

As long as different streams of dark matter do not cross each other,
the cross-section for their collision does not enter the equations.
Therefore the perturbative framework, which is the basis for the
recently suggested renormalization methods, is insensitive to dark
matter collisional properties. Even though in the minimal dark
matter model, the particles have a negligible self-interaction
cross-section, one may also envisage an opposite regime, where the
collisions have a huge cross-section, and are maximally inelastic.
We dub this particular model for dark matter as {\it sticky dark
matter}.

Of course, sticky dark matter is a terrible candidate for the
cosmological dark matter. This is because, as a result of frequent
and inelastic collisions, all the dark matter particles would sink
into the center of a halo, probably forming a black hole, or a small
compact disk (if prevented by the initial angular momentum of the
halo). However, it provides an interesting theoretical test study to
examine the non-perturbative effect of shell-crossing, as sticky and
collisionless dark matter obey the exact same equations within the
(single-stream) perturbation theory.

In order to make this comparison, we resort to the simplest version
of the halo model \cite{Cooray:2002di}, where dark matter haloes
cluster according to the linear matter power spectrum with a
constant bias, which only depends on their mass. The structure
within a halo of collisionless dark matter is well approximated by
an NFW form \cite{Navarro:1996gj}: \beq \rho(r)=
\frac{\rho_s}{(r/r_s)(1+r/r_s)^2}, \eeq which extends out to
$r_{vir} = c r_s$, where the concentration parameter, $c$, as a
function of the total enclosed mass is also constrained from
simulations.

It is known that a combination of halo-halo correlations, plus
single halo auto-correlation terms gives a reasonable approximation
to the dark matter non-linear power spectrum in numerical
simulations (e.g. \cite{Smith:2002dz}). With this qualitative
picture at hand, we can thus examine the case of sticky dark matter,
by collapsing each NFW halo into a point.

Fig. (\ref{deltak}) compares the dimenionless power spectra,
$\Delta^2(k) = \frac{k^3 P(k)}{2\pi^2}$, for collisionless and
sticky dark matter models, in the context of the halo model. Here,
we assume the WMAP 1st year best fit concordance cosmological
parameters \cite{Spergel:2003cb}. We see that the difference between
the two models, which characterizes the level of non-perturbative
effects in the dark matter power spectrum is $\sim 1\%$ at $k \sim
0.1 \mpc^{-1}$, which is the region relevant for probing baryonic
acoustic oscillations. This level rises to order unity at $k \sim 1
\mpc^{-1}$, indicating the complete breakdown of (renormalized)
perturbation theory, or any single-stream renormalization group
method.

\begin{figure}
\includegraphics[width=0.9\linewidth]{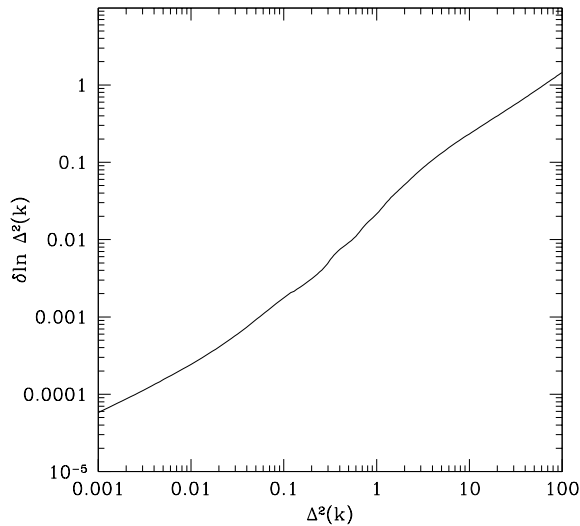}
\caption{\label{deltadelta} The level of non-perturbative effects in
the non-linear power spectrum (see Fig.\ref{deltak}), as a function
of non-linear power. }
\end{figure}

Fig. (\ref{deltadelta}) shows the same difference between the sticky
and collisionless power spectra, as a function of the non-linear
power. It is interesting to notice that non-perturbative effects do
not dominate the power until $\Delta^2(k) \sim 100$, which is far
into the non-linear regime. This may explain why renormalization
methods based on the perturbation theory, such as
\cite{Crocce:2005xy,McDonald:2006hf}, can do so well in the
quasilinear regime.

A question that may arise is if we are stretching the halo model
beyond its level of applicability. It is true that the halo model
has had significant phenomenological success in fitting simulations
and observations (e.g. \cite{Smith:2002dz} and
\cite{Zehavi:2003ta}). However, it suffers from unphysical features,
such as spurious power on very large scales, which is caused by the
presence of the 1-halo term \cite{Scoccimarro:2000gm,
Cooray:2002di}. However, the difference shown in Figs.
(\ref{deltak}) and (\ref{deltadelta}) is dominated by the 2-halo
term for $k \lesssim 0.3 \mpc^{-1}$ (at least down to near-horizon
scales). Even without any reference to the halo model, it is clear
that the large scale correlation function is, to some extent,
affected by the change in the profile of the virialized regions, say
from an NFW (for collisionless dark matter) to a point-like profile
(for sticky dark matter). The difference in the 2-halo term, in the
context of the halo model, gives an approximation for the magnitude
of this effect.

As vorticity cannot be locally generated in the single stream regime
of dark matter collapse, a different test for the breakdown of the
single-stream  perturbation theory is when the vorticity and the
divergence of the velocity field become comparable
\cite{Scoccimarro:2000zr}. This happens at $ k \sim 2 \mpc^{-1}$,
which is consistent with our results. However, in contrast to our
approach, it is not clear how the vorticity to divergence ratio is
related to the level of non-perturbative effects in the power
spectrum within the quasilinear regime.

What if a single-stream perturbative method agrees with numerical
N-body simulations at a better level than predicted by our simple
exercise? Then I would argue that this level of accuracy is not
warranted for any single-stream calculation, and thus any such
agreement is {\it accidental}, unless it is justified based on a
non-perturbative method.

Finally, we should point out that sticky dark matter is very similar
to the so-called adhesion model
\cite{1992ApJ...393..437K,1994ApJ...428...28M}, that has been
introduced in the context of Zel'dovich approximation \footnote{I
thank Roya Mohayaee for pointing this out.}. Adhesion model, which
is realized as the zero/small viscosity limit of a (single stream)
fluid, has been introduced to enable the use of analytic Zel'dovich
approximation beyond shell-crossing (or pancake formation). However,
direct comparison of the adhesion model to the sticky dark matter
might be misleading, as it mixes the perturbative terms, missing in
the Zel'dovich approximation, with the non-perturbative adhesive
feature.

In conclusion, in this note, we have pointed out that any
renormalized perturbation theory method is ultimately limited to the
context of single stream fluid equations, and thus does not capture
the non-perturbative shell-crossing events. To estimate the level of
non-perturbative effects, we introduced a toy model of sticky dark
matter, which only differs from collisionless dark matter in the
shell-crossing regime. This provides us with a framework to estimate
the level of accuracy expected from renormalized perturbation
theory, showing that non-perturbative effects could be $\sim 1\%$ at
scale of baryonic acoustic oscillations, but do not dominate until
deep in the non-linear regime.

%

I would like to thank Simon DeDeo, Roya Mohayaee, Roman Scoccimarro,
Anze Solsar, and especially Patrick McDonald for helpful comments
and stimulating discussions. I also thank the organizers of the
Benasque Cosmology Workshop, for providing the atmosphere that
inspired this work.


\bibliographystyle{utphys_na}
\bibliography{sticky}

\end{document}